\documentclass{appolb}

\usepackage{graphicx}
\usepackage{amssymb}
\usepackage{hyperref}

\begin{document}

\title{Ultra-relativistic light-heavy nuclear collisions \\ and collectivity%
\thanks{Presented by WB at Excited QCD 2015, Tatranska Lomnica, 8-14 March 2015}}
\author{Wojciech Broniowski$^{1,2}$, Piotr Bo\.zek$^3$, Maciej Rybczy\'nski$^1$, Enrique Ruiz Arriola$^4$
\address{$^1$Institute of Physics, Jan Kochanowski University, 25-406~Kielce, Poland \\
               $^2$The H. Niewodnicza\'nski Institute of Nuclear Physics, \\ Polish Academy of Sciences, 31-342 Cracow, Poland\\
               $^3$AGH University of Science and Technology, Faculty of Physics and Applied Computer Science, 30-059 Krakow, Poland\\
               $^4$Departamento de F\'{\i}sica At\'{o}mica, Molecular y Nuclear and Instituto Carlos I \\ de  F{\'\i}sica Te\'orica y Computacional, 
 Universidad de Granada, \\ E-18071 Granada, Spain
               }
}
\maketitle
\begin{abstract}
We briefly review highlights for ultra-relativistic light-heavy collisions (\mbox{p-Pb}, d-Au, $^3$He-Au, $^{12}$C-Au) which display collective evolution, with the same very
characteristic features as in the A-A systems.
\end{abstract}
\PACS{25.75.-q,25.75.Dw,25.75.Nq}
  
\bigskip \bigskip
  
This talk is based on Refs.~\cite{Bozek:2012gr,Bozek:2013df,Bozek:2013uha,Bozek:2013ska,Bozek:2014cya,Bozek:2014cva,Bozek:2015qpa,Broniowski:2013dia}, 
where more details and complete references may be found.
  
Collectivity of the evolution in the intermediate phase of ultra-relativistic nuclear reactions leads to specific, very characteristic signatures. Due to very large density of the initial 
fireball, collective flow of the medium is generated, which determines many features found in experiments as well as in modeling based on 
hydrodynamics of transport models. The most vivid hallmarks are:

\begin{enumerate}
 \item The ridge structure in two-particle correlations in relative azimuth and pseudorapidity. In particular, the collimation of flow at distant 
 pseudorapidities yields the away-side ridge. Observation of this phenomenon in p-Pb collisions, as well as in the highest multiplicity \mbox{p-p} collisions, have changed 
 our view on the dynamics of light-heavy reactions~\cite{Khachatryan:2010gv,Adare:2013piz,Aad:2012gla,Chatrchyan:2013nka,Abelev:2012ola}.  
 \item Mass ordering of such observables as the mean transverse momentum or the harmonic flow coefficients. Due to emission from fluid elements
 moving with large collective velocity, heavier hadrons acquire 
 more momentum than the lighter ones. In particular, collective modeling of p-Pb collisions~\cite{Bozek:2013ska} reproduces the data seen in proton-nucleus collisions~\cite{Abelev:2013bla}.
 \item The near-equality of higher-order cumulants (involving 4, 6, 8, etc., particles) for the harmonic flow coefficients. The phenomenon is caused by the collective nature of correlations, and holds also for \mbox{p-Pb} 
 reactions~\cite{Bzdak:2013rya}.
 \item The fall-off of the HBT correlation radii with the transverse momentum of the pair also indicates flow in heavy-light systems~\cite{Bozek:2013df,Bozek:2015qpa}.
 It has recently been observed in p-Pb collisions~\cite{Adam:2015ewa}.
 \item Transverse-momentum fluctuations, as generated by the mechanism of Ref.~\cite{Broniowski:2009fm}.
 \item Long-range event-plane correlations in pseudorapidity and the torque effect~\cite{Bozek:2010vz,Jia:2014ysa,Bozek:2015bna}.
\end{enumerate}

\begin{figure}[tb]
\centerline{\includegraphics[angle=0,width=.65 \textwidth]{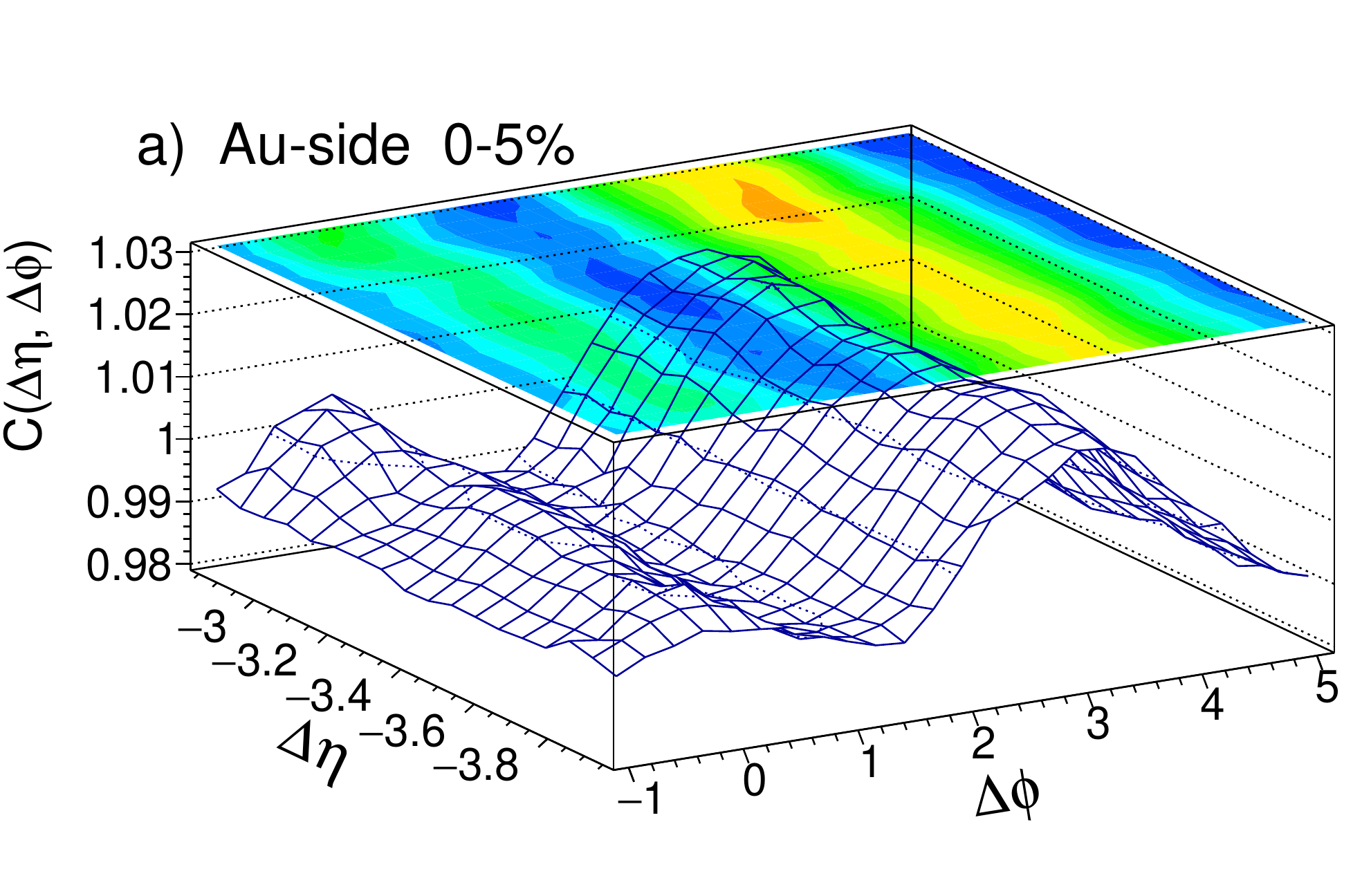}}
\vspace{-9mm}
\centerline{\includegraphics[angle=0,width=.65 \textwidth]{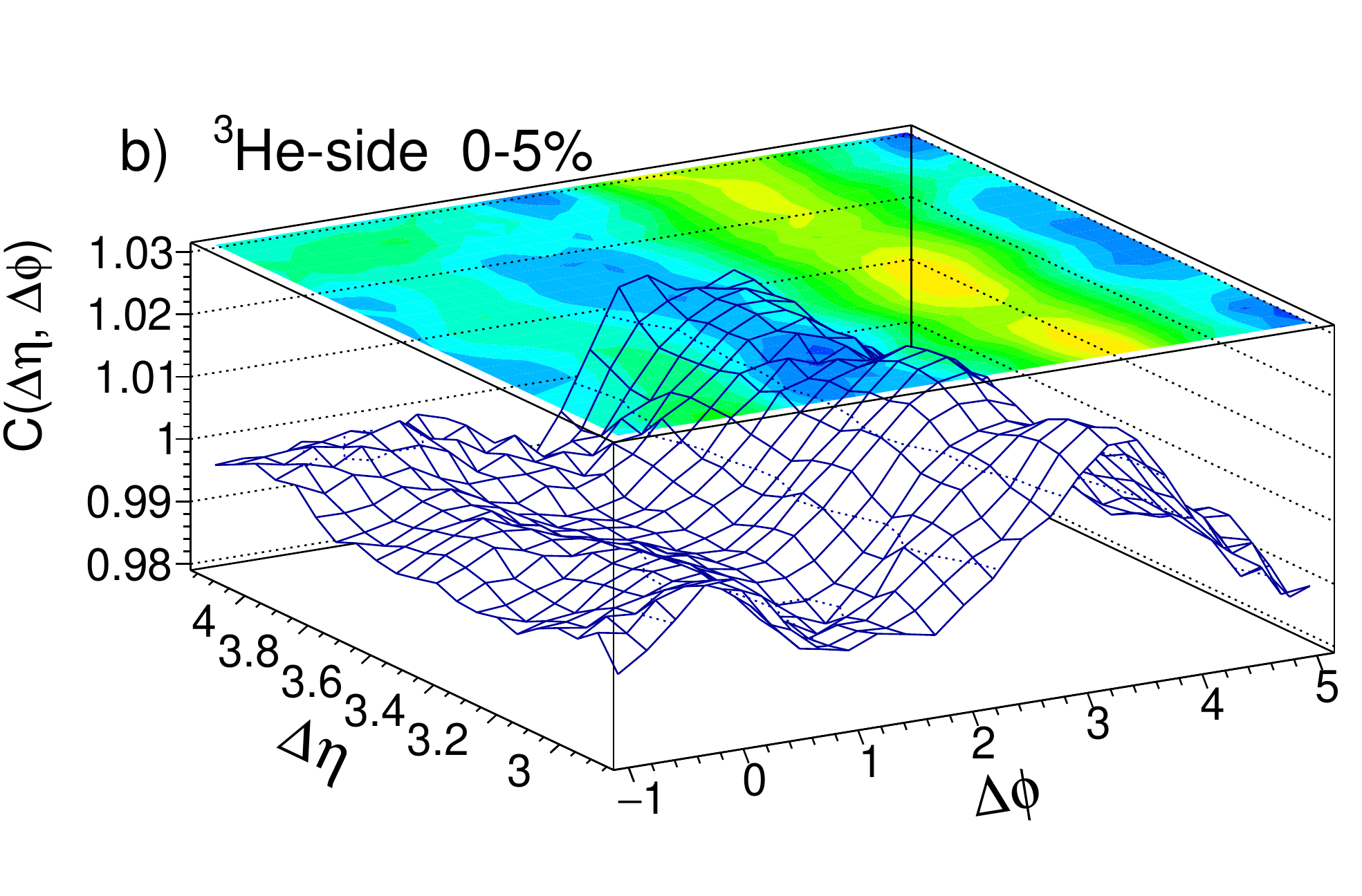}}
\vspace{-2mm}
\caption{The ridge effect in $^3$He-Au collisions, seen in the two-particle correlation function in relative azimuth and pseudorapidity. 
(taken from Ref.~\cite{Bozek:2015qpa}). \label{fig:He}}
\end{figure}

Effects which involve harmonic flow rely on the shape-flow transmutation, thus are sensitive to the modeling of the initial state and fluctuations therein. Investigations of 
small systems serve to set the limits on applicability of hydrodynamics or transport theory, and differentiation with other approaches, for instance those  
based on the QCD saturation phenomena~\cite{Dusling:2012cg}.

\begin{figure}[tb]
\centerline{\includegraphics[angle=0,width=.65 \textwidth]{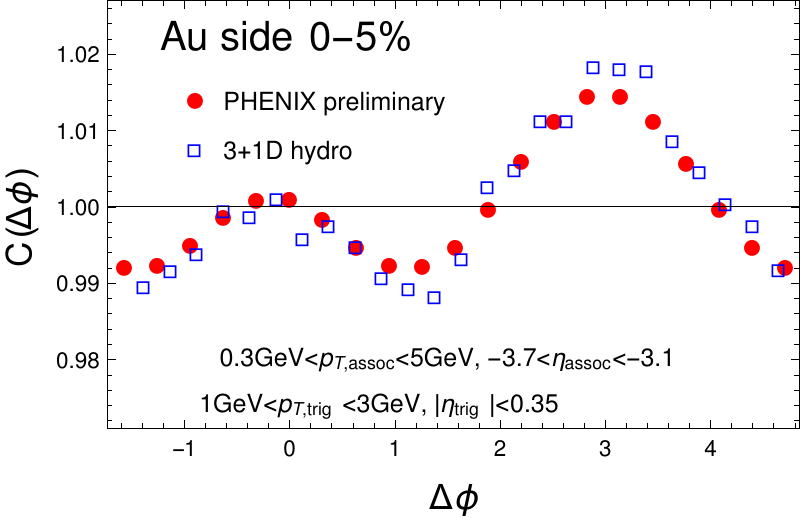}} \vspace{-6mm}
\centerline{\includegraphics[angle=0,width=.65 \textwidth]{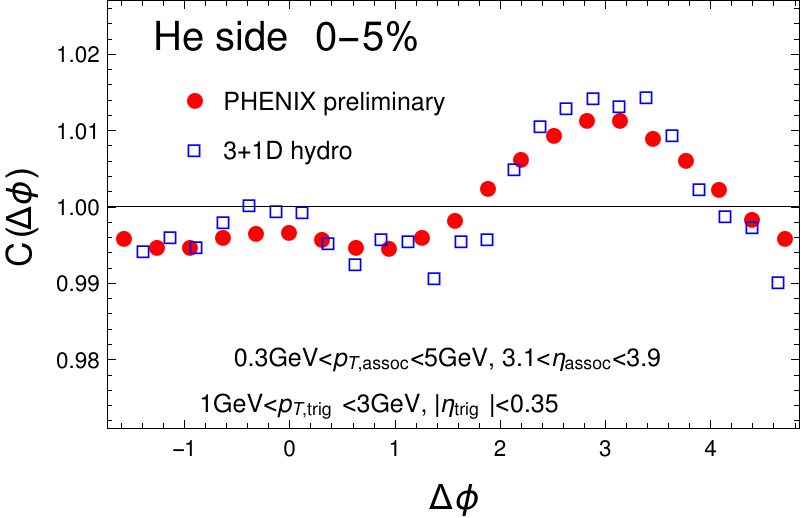}}
\caption{Projected two-particle correlation function in relative azimuth. The PHENIX data come from Ref.~\cite{phenixnapa}. 
(taken from Ref.~\cite{Bozek:2015qpa}). \label{fig:Hep}}
\end{figure}

Results for the p-A and d-A~\cite{Bozek:2011if} systems have been extensively presented in the cited literature, hence we do not discuss them here. In Figs.~\ref{fig:He} and~\ref{fig:Hep} we show an outcome of a recent 
study~\cite{Bozek:2015qpa} for $^3$He-Au collisions, plotting the correlation function
\begin{eqnarray}
C(\Delta\eta,\Delta \phi) &=& \frac{S(\Delta \eta, \Delta \phi)}{B(\Delta \eta, \Delta \phi)}, \label{eq:C}
\end{eqnarray}
where the signal $S$ is constructed from pairs of particles with the relative pseudorapidity $\Delta \eta$ and the 
relative azimuth $\Delta \phi$, while the background $B$ is evaluated with the mixed events. 
The kinematic cuts indicated in the figure correspond to the PHENIX experiment~\cite{phenixnapa}. We note the formation of the ridges, both on the Au and $^3$He sides, clearly indicating the 
collectivity of the dynamics. We note that the model based on hydrodynamics is in very good agreement with the data, as can be seen from Fig.~\ref{fig:Hep}.

\begin{figure}[tb]
\centerline{\includegraphics[angle=0,width=.715 \textwidth]{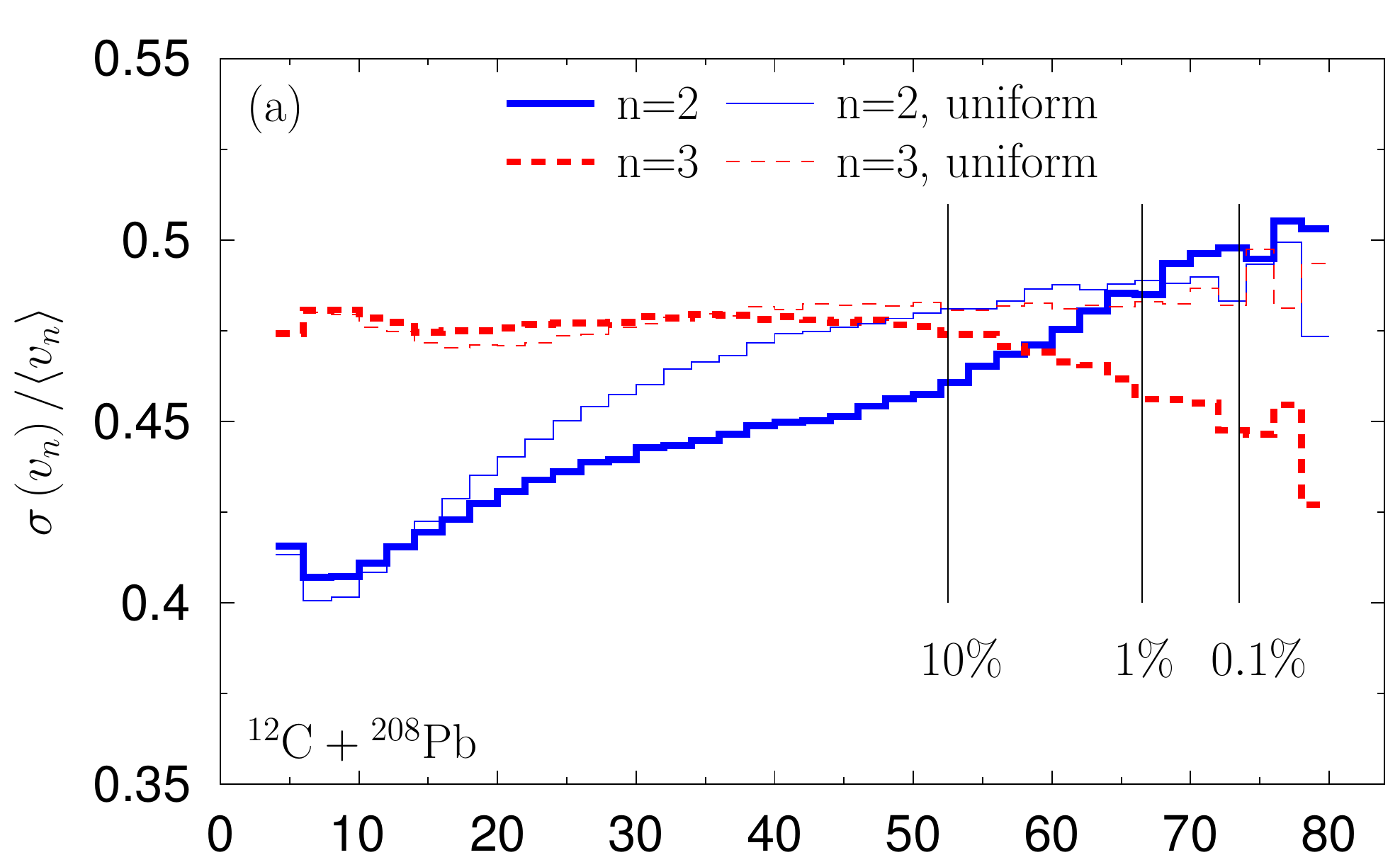}}
\vspace{-4mm}
\centerline{~ \includegraphics[angle=0,width=.7 \textwidth]{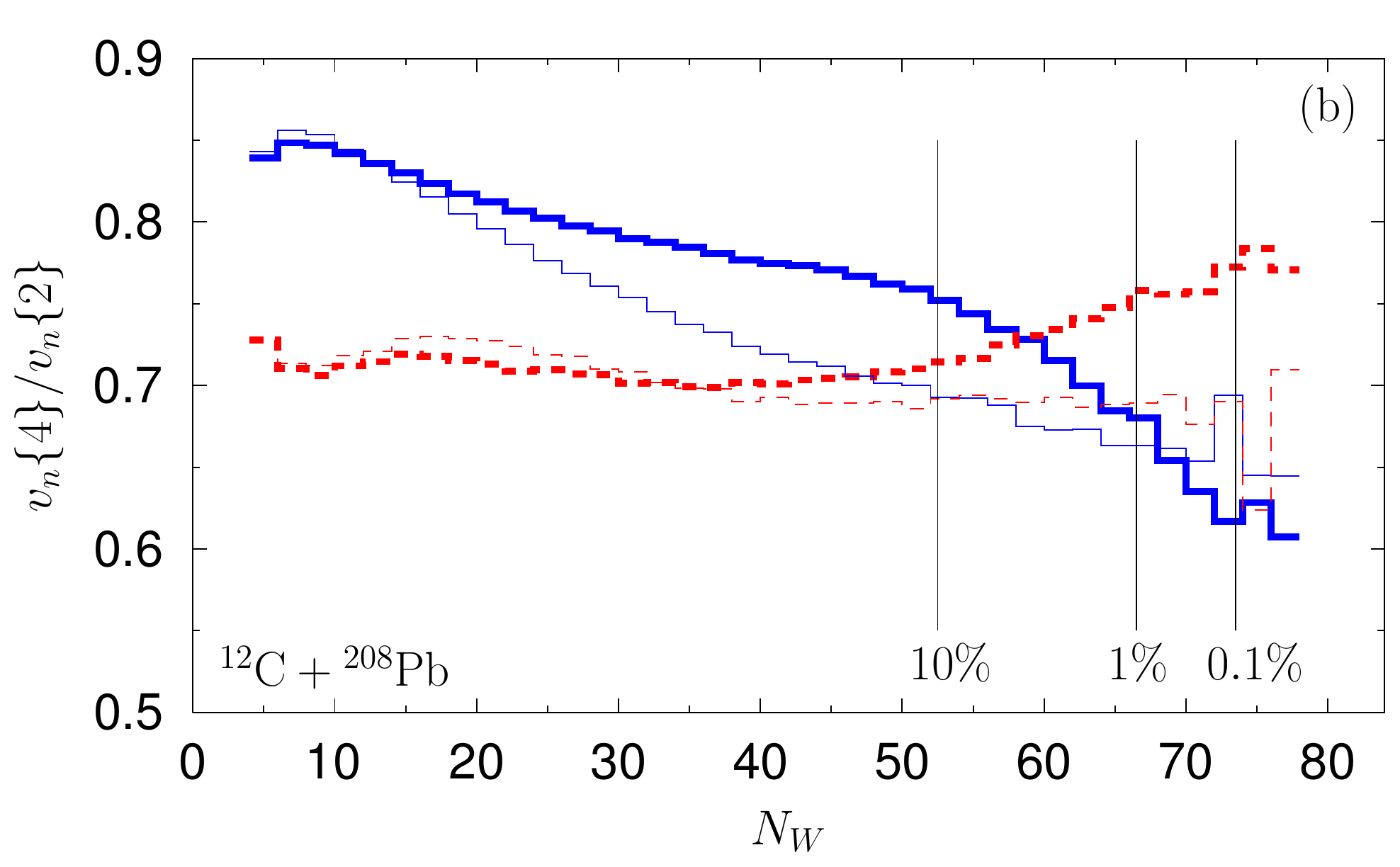}}
\vspace{-2mm}
\caption{Ratios~(\ref{eq:sc}) vs the number of wounded nucleons for the $^{12}$C-$^{208}$Pb collisions computed from the mixed Glauber model simulations~\cite{Rybczynski:2013yba} at the SPS energies, with
the nucleon-nucleon inelastic cross section  $\sigma^{\rm inel}_{NN}=32$~mb. Centralities are indicated by vertical lines. \label{fig:C}}
\end{figure}

The shape-flow transmutation has, to a good accuracy, the approximate feature that the distribution of scaled (i.e, divided by the average) eccentricities in the initial
state is equal to the distribution of the scaled harmonic flow~\cite{Bozek:2014cya}. The property holds as long as the response of the system to small eccentric perturbations is linear.
Then one finds corresponding equalities for scaled statistical event-by-event measures for $n=2, 3$ (higher-order harmonics have nonlinear response)
\begin{eqnarray}
\frac{\sigma(\epsilon_n)}{\langle \epsilon_n \rangle} = \frac{\sigma(v_n)}{\langle v_n \rangle}, \;\;\; 
\frac{\epsilon_n\{4\}}{ \epsilon_n\{2\} } =  \frac{v_n\{4\}}{v_n\{2\}}, \;\;\; {\rm etc.}, \label{eq:sc}
\end{eqnarray}
where $\{m\}$ indicates quantities obtained form $m$-particle cumulants. Formulas~(\ref{eq:sc}) have important practical significance, as they allow for making predictions for the measurable
flow coefficients solely by modeling the eccentricities in the initial state, without the costly hydrodynamic simulations and hadronization. 

In Ref.~\cite{Broniowski:2013dia} a new methodology of studying the ground-state correlations in nuclear distributions has been proposed. It is based on the shape-flow transmutation, which 
carries over the initial eccentricities to the flow coefficients. As an interesting example, the $^{12}$C nucleus, due to strong $\alpha$ clusterization, may be viewed as a small triangle. A collision at ultra-relativistic energy 
proceeds in a time much shorter from any characteristic nuclear time scale, hence a frozen ground-state configuration is seen. The collision forms a triangular fireball, which upon evolution leads to 
increased triangular flow. The picture is blurred to some extent with the fluctuations and averaging over orientations, nevertheless a substantial effect persists.

In Fig.~\ref{fig:C} we show the Glauber model predictions for the  $^{12}$C-$^{208}$Pb collisions. We compare the clustered wave function (thick lines) to uniform distribution (thin lines). We note large effects, especially 
at low centralities. We note that the curves for the triangular flow coefficients change character at $c\sim10\%$, where the clustered and uniform cases depart from each other: the scaled 
standard deviation of Fig.~\ref{fig:C}(a) decreases 
with $N_W$ for the clustered case, whereas for the uniform case it remains almost constant. The origin of this behavior is geometric. As $N_W$ increases, the $^{12}$C triangle is oriented more and more face-on with respect 
to the reaction plane, hence average triangularity increases and the ratio  ${\sigma(\epsilon_n)}/{\langle \epsilon_n \rangle}$ decreases. The behavior for the ellipticity is opposite. 
Similarly, the cumulant ratios of Fig.~\ref{fig:C}(b) change behavior around $c=10\%$. These results, showing 
qualitative and quantitative sensitivity of the harmonic flow to specific features of the ground-state wave function, prove the feasibility of the 
proposed new method of studying low-energy nuclear structure with techniques developed for ultra-relativistic heavy-ion collisions. 

\bigskip
                               
Research supported by the
National Science Center grants DEC-2012/\-05/\-B/\-ST2/02528 and DEC-2012/06/A/ST2/00390, 
by the Polish Ministry of Science and Higher Education (MNiSW), by PL-Grid Infrastructure, and by Spanish
DGI (grant FIS2014-59386-P).

\bigskip

\bibliography{hydr}

\end{document}